\documentclass[aps,prmaterials,superscriptaddress,reprint]{revtex4-2}

\usepackage{amsmath,amsfonts,amssymb,bm}
\usepackage{graphicx}
\usepackage{hyperref}
\usepackage{xcolor}

\begin{document}

\title{Strain effects in phosphorous bound exciton transitions in silicon}

\author{Teemu Loippo}
\affiliation{Department of Physics and Nanoscience Center, University of Jyv{\"a}skyl{\"a}, P.O. Box 35, FI-40014 University of Jyv{\"a}skyl{\"a}, Finland}

\author{Antti Kanniainen}
\affiliation{Department of Physics and Nanoscience Center, University of Jyv{\"a}skyl{\"a}, P.O. Box 35, FI-40014 University of Jyv{\"a}skyl{\"a}, Finland}

\author{Juha T. Muhonen}
\email{juha.t.muhonen@jyu.fi}
\affiliation{Department of Physics and Nanoscience Center, University of Jyv{\"a}skyl{\"a}, P.O. Box 35, FI-40014 University of Jyv{\"a}skyl{\"a}, Finland}

\begin{abstract}
Donor spin states in silicon are a promising candidate for quantum information processing. One possible donor spin readout mechanism is the bound exciton transition that can be excited optically and creates an electrical signal when it decays. This transition has been extensively studied in bulk, but in order to scale towards localized spin readout, microfabricated structures are needed for detection. As these electrodes will inevitably cause strain in the silicon lattice, it will be crucial to understand how strain affects the exciton transitions. 
Here we study the phosphorous donor bound exciton transitions in silicon using hybrid electro-optical readout with microfabricated electrodes. We observe a significant zero-field splitting as well mixing of the hole states due to strain. 
We can model these effects assuming the known asymmetry of the hole g-factors and the Pikus-Bir Hamiltonian describing the strain. In addition, we describe the temperature, laser power and light polarization dependence of the transitions. Importantly, the hole-mixing should not prevent donor electron spin readout and using our measured parameters and numerical simulations we anticipate that hybrid spin readout in a silicon-on-insulator platform should be possible, allowing integration to silicon photonics platforms.
\end{abstract}

\date{\today}

\maketitle

\section{Introduction}
\label{sec:intro}
Donor spin qubits in silicon are a promising candidate for processing quantum information \cite{Kane1998,Morton2008,Morello2010,Tyryshkin2012,Pla2012,Steger2012,Kamyar2013,Muhonen2014,He2019,Madzik2022}. Their application potential is however somewhat constrained by the lack of an optical interface, and considerable attention has been recently focused on spin-photon interfaces in silicon \cite{Songtao2020, Hollenbach2020, Bergeron2020, Redjem2020, Durand2021}. One possible readout avenue is the donor bound exciton transition \cite{Kaminskil1980, Karaiskaj2001, Yang2006, Steger2011, Gullans2015}, which however decays mostly via Auger recombination hence not enabling an efficient optical readout protocol. Hybrid electro-optical readout, where the spin-selective transition is excited optically but readout electrically, is possible and has been demonstrated with both direct electrical and capacitive readout \cite{Ross2019} in bulk. Scaling the electro-optical readout towards single-spin level will, however, require moving to microfabricated structures as well as most likely taking advantage of silicon photonics platforms. So far, only one demonstration using microfabricated electrodes exists \cite{Lo2015} and the integration to photonics platforms remains an open challenge. The latter will require moving to silicon-on-insulator (SOI) substrates where photoluminescence measurements have shown the existence of the bound exciton transition \cite{Sumikura2011} but no electrical readout has been demonstrated.

One open question when moving towards microfabricated structures on SOI substrates is the role of strain in the exciton readout. At low temperatures the different thermal expansion coefficients of silicon and metal cause sharp strain profiles around the electrodes. This might be especially important when all the spins are located in the surface layer in an SOI structure (although see Section \ref{sec:discussion} for further discussion on this point). Several recent studies \cite{PLa2018, Mansir2018, Ranjan2021} have highlighted how strain can substantially change the donor spin states and their decoherence properties. Substantially less attention in this context has so far been paid to strain effects of the hole and exciton states.

Here we report experiments demonstrating the electro-optical donor exciton response using microfabricated electrodes and focused light on doped natural silicon. We have performed experiments on both silicon grown with the Czochralski method (CZ) and float-zone method (FZ). All data shown is from the FZ silicon unless otherwise mentioned. The strain from the on-chip electrodes creates significant effects but importantly does not create any fundamental obstacles for localized spin readout. We found strain-induced heavy-hole light-hole splitting to be significant in all samples and we see evidence of avoided crossings between the heavy-hole and light-hole transitions indicating a coupling term in the Hamiltonian. We can model these effects with remarkable agreement assuming the known asymmetry of the hole g-factors and the Pikus-Bir Hamiltonian describing the strain, but we also have unresolved questions regarding temperature dependence and light polarization response which is complicated by the mixing of the different hole states. Nevertheless, using the extracted parameters we predict with numerical simulations how the exciton signal should behave in an SOI structure where the strain effects must be considered with care.

\section{Experimental setup}
\label{sec:exp}
\begin{figure}
\includegraphics[width=0.45\textwidth]{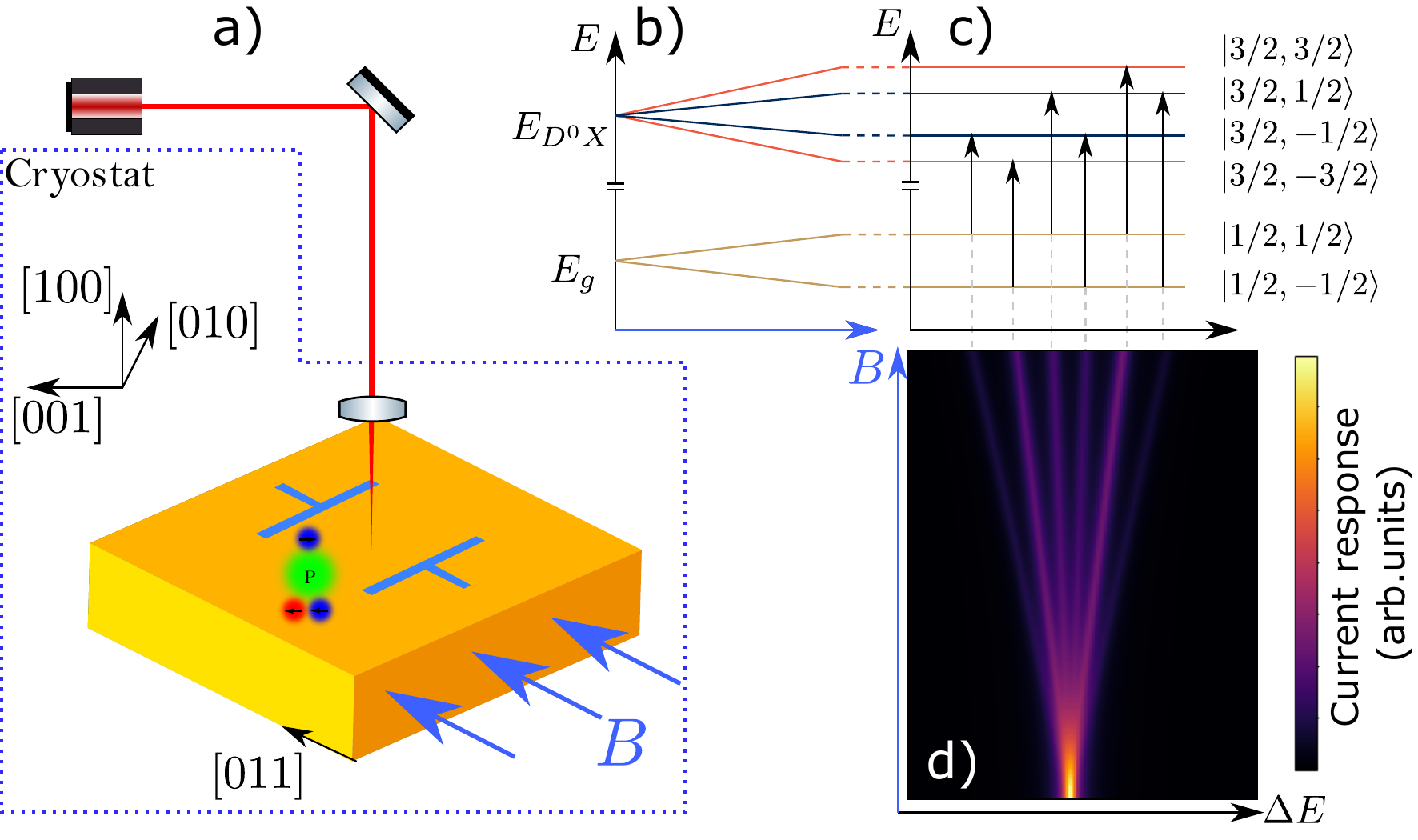}
\caption{a) A schematic of our sample, the measurement geometry and the optical setup. Light is guided to the sample in free space via a fixed lens. The magnetic field is in-plane and parallel to the direction of measured current and the [011] crystal direction. b) Energy level diagram showing the Zeeman splitting of donor electron ($E_g$) and hole ($E_{D^0X}$) states. c) Arrows showing the six allowed exciton transitions in a non-zero magnetic field. d) Simulated current signal of the exciton transitions as a function of magnetic field in strain-free silicon. Transition brightness corresponds to the Clebsch-Gordan coefficient of the transition.}
\label{fig:exciton_scheme}
\end{figure}

At low temperatures phosphorous donors in silicon lattice can bind an electron around them in a Coulomb potential forming a system that resembles a hydrogen atom but with a Bohr radius of around 1.8~nm \cite{Smith2017}. The spin of this bound electron, in combination with the phosphorous nuclear spin, has been shown to be a promising quantum computing unit \cite{Kane1998,Pla2012}. Unfortunately for optical applications, the loose binding potential means that the "atomic" transitions of this system are in the technologically inconvenient terahertz regime. There is nevertheless a bound exciton transition at a more technologically convenient 1078~nm wavelength. The electron of the exciton will form a singlet pair with the donor electron, leaving the hole 3/2-spin as the spin degree of freedom in the bound exciton state. Hence, there are six allowed transitions between the original donor spin state and the exciton state in a finite magnetic field, these are depicted in Fig.~\ref{fig:exciton_scheme}(b-c). We neglect the hyperfine coupling between the donor electron spin and the nuclear spin throughout the paper as our linewidths are not narrow enough to resolve it. This could be ameliorated by moving to isotopically pure silicon \cite{Karaiskaj2001,Steger2012}. As any single transition originates from a particular donor electron spin state, the existence or non-existence of these transitions can be used for electron spin readout. The decay of the exciton happens mostly via Auger recombination, ending up with an ionized donor and an extra "hot" electron now in the conduction band. This causes a change in the conductivity of the silicon, which we detect.

The measurement scheme is presented in Fig.~\ref{fig:exciton_scheme}(a). The samples studied are uniformly doped silicon chips onto which two gold electrodes with length of 100~$\mu$m and separated by 60~$\mu$m (CZ sample) or 100~$\mu$m (FZ sample) are fabricated using electron beam lithography and liftoff techniques.
Experiments were carried out inside a ``dry'' dilution fridge, which however is mostly operated at 3.3~K temperatures for the data presented here. The laser spot diameter is roughly 50~$\mu$m in the data presented, and hence far from the diffraction limit. Sample is placed on a piezo stage which allows us to align and focus the laser spot with the readout electrodes using an infrared camera outside the cryostat. The electrodes connect to a voltage (CZ measurements) or a current source (FZ measurements) and measured signals are amplified using low noise amplifiers. Current-voltage characteristics of the sample are presented in Supplementary Material section~\ref{sec:IVdetails}. We have LEDs (peak wavelength 1020~nm) inside the cryostat to provide light with energy above the silicon bandgap. The magnetic field direction is perpendicular to the light propagation direction, meaning we are using the Voigt geometry here, allowing us to address all the transitions with linearly polarized light. The magnet is calibrated with a Hall sensor. We have a small uncertainty in the location of the sample with regards to the calibration location, from which we estimate a conservative error range of $\pm 5$~\% for the magnetic field calibration. See Supplementary Material section~\ref{sec:measurementsetup} for more details about the measurement setup.

Figure \ref{fig:exciton_scheme}(b-c) shows schematically the six allowed bound exciton transitions and Fig.\ref{fig:exciton_scheme}(d) depicts the signal we would expect to get in different magnetic fields. The actual measured signal is shown in Fig. \ref{fig:4K_magnet_sweep}(c). There are several main differences which we assign to the effects of strain.

\section{Magnetic field dependency and strain effects}
\label{sec:mag}
As explained above, the transitions we observe happen essentially between a single-electron spin-state (D$_0$) and a hole spin-state (D$_0$X). Hence, we need to write down the Hamiltonians for both the originating and the final state of the system, while taking strain effects into consideration. We note that we do not try to model the center frequency of the transition but rather only the magnetic field dependency. Our center transition frequency at zero field (1078.180 $\pm$ 0.015~nm) is close to what has been reported before \cite{Safonov1996,Yang2006,Yang2009}, we assign the slight discrepancy to the substrate material that is not especially free of impurities.

The magnetic field dependency of the neutral donor state is assumed to simply follow
\begin{equation}
    H_D^Z=\frac{1}{2}\mu_B g_d \mathbf{B} \cdot \bm{\sigma},
    \label{eq:donorZeeman}
\end{equation}
where $\mu_B$ is the Bohr magneton, $g_d=1.9985$ is the donor electron $g$-factor, $\mathbf{B}$ is the magnetic field vector and $\bm{\sigma}$ is a vector containing the Pauli spin matrices. We assume isotropic $g$-factor for the donor electron, although the isotropy might be broken by strain \cite{Wilson1961,Mansir2018}. Nevertheless, the expected anisotropy is of the order of 10$^{-3}$\cite{Usman2018}, much smaller than the strain effects for the holes as we will show below, and can be neglected here. We also neglect the hyperfine interaction since we cannot resolve it. There are also strain variation in the electron energy levels coming from the variation of the hyperfine interaction as a function of strain but also these are expected to be small \cite{Wilson1961,PLa2018, Mansir2018, Ranjan2021} compared to the effects we study here.

For the magnetic field dependence of the hole we use the anisotropic $g$-factor model \cite{Bir1963} 
\begin{equation}
    H^Z_B = \mu_B\left( g_1\mathbf{J}\cdot \mathbf{B} + g_2\mathbf{J}^3 \cdot \mathbf{B} \right),
    \label{eq:holezeeman}
\end{equation}
where $\mathbf{J}$ is a vector containing 3/2-spin matrices in $x, \: y \: \text{ and } z $ directions (which we take to correspond to the crystal directions of [100] and equivalent in order to match the strain directions in Eq.~(\ref{eq:HPB}) below), $\mathbf{B}$ is the magnetic field vector, and $g_1$ and $g_2$ are the isotropic and anisotropic $g$-factors, respectively.

\begin{figure}
\includegraphics[width=0.45\textwidth]{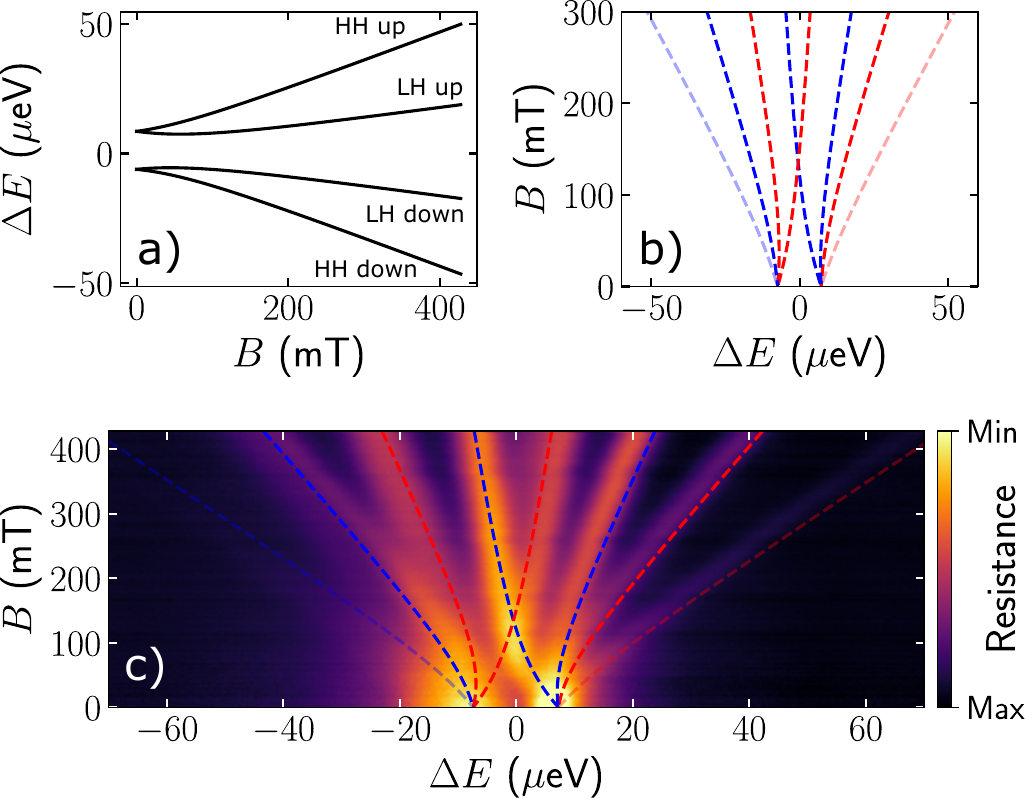}
\caption{a) Strain split hole eigenenergies and b) exciton transition energies calculated using Eqs.~(\ref{eq:donorZeeman}), (\ref{eq:holezeeman}) and (\ref{eq:HPB}). $B$ is the applied external magnetic field. The magnetic field lies in the [011] crystal direction. The labels show the dominating state at high magnetic fields.
c) Measured transition energies as a function of the magnetic field, in the [011] direction, overlaid with the calculated transition energies presented in (b). The center transition frequency is adjusted to match the data. Blue lines indicate transitions involving the donor electron spin up state and red lines transitions coming from donor electron spin down state. Transitions that are forbidden without the hole mixing are shown with transparent lines.}
\label{fig:4K_magnet_sweep}
\end{figure}

Neglecting the strain effects, that is, only using Hamiltonians described by Eq.~(\ref{eq:donorZeeman}) for the donor and Eq.~(\ref{eq:holezeeman}) for the hole, we will get exciton transition lines shown in Fig. \ref{fig:exciton_scheme}(d). Comparing this to our actual measurement data shown in Fig. \ref{fig:4K_magnet_sweep}(c), reveals several features not described by the model. The most prominent discrepancy is the existence of two peaks at zero magnetic field. We assign this zero field splitting (ZFS) to the splitting of the heavy-hole and light-hole states caused by strain. The strain effects affecting hole states in silicon are conventionally modeled with the Pikus-Bir strain Hamiltonian \cite{bir1974}
\begin{align}
    H_{PB}(\bm{\epsilon}) = &a\mathrm{Tr}(\bm{\epsilon}) + b \sum_{i=x,y,z} \left( J^2_i - \frac{\mathbf{J}^2}{3} \right) \epsilon_{ii} +\nonumber \\
    &\frac{d}{\sqrt{3}} \sum_{i\neq j}(J_iJ_j + J_jJ_i)\epsilon_{ij}, 
    \label{eq:HPB}
\end{align}
where $\bm{\epsilon}$ is the strain matrix and $a,b$ and $d$ are deformation potentials. The parameter $a$ only changes the center frequency of the transition and is not important for our model here. The other factors $b$ and $d$ we use as fit parameters. The extracted values are listed in Table \ref{tab:fits}. For strain, we use values from a \textsc{COMSOL} simulation of the strain caused by the different thermal expansion coefficients of silicon and the metallic electrodes and the stage, see section \ref{sec:strain} for details. Note that as the strain and parameters $b$ and $d$ are always multiplied, any discrepancy in the strain parameters will then directly affect these values.

\begin{table}
     \centering
     \caption{Extracted values for deformation potentials $b$ and $d$ along with hole g-factors $g_1$ and $g_2$ from fitting Eqs.~(\ref{eq:HPB}) and (\ref{eq:holezeeman}) to the data shown in Fig.\ref{fig:4K_magnet_sweep}(c). Reference values for $b$ and $d$ are taken from \cite{Blacha1984}. Reference values for $g_1$ and $g_2$ in [100] and [111] orientations are from \cite{Kaminskil1980}. $g_{LH}$ and $g_{HH}$ were calculated using Eqs.~(\ref{eq:EHH}--\ref{eq:gHH}). The reference values in [011] orientation are taken from \cite{Litvinenko2016}.}
     \begin{tabular}{c c c c c c c}
     \hline\hline
           & $b$ (eV) & $d$ (eV) & $g_1$ & $g_2$ & $g_{LH}$ & $g_{HH}$\\
         \hline
          
          This work [011] & -7 & -4 & 0.83 & 0.22 & 1.40 & 1.28 \\
          
          Reference & -2.2 & -5.1 & - & - & - & - \\
          
          Reference [100] & - & - & 0.8 & 0.24 & 0.86 & 1.34 \\

          Reference [111] & - & - & 0.86 & 0.21 & 1.57 & 1.27 \\
          
          Reference [011] & - & - & 0.83 & 0.225 & 1.409 & 1.285 \\
          \hline\hline
     \end{tabular}
     \label{tab:fits}
 \end{table}

By itself the Pikus-Bir strain Hamiltonian simply splits the states at zero-field. However, the anisotropic component of the Zeeman Hamiltonian ($g_2$ in Eq.~(\ref{eq:holezeeman})) causes coupling of the hole states and we get the avoided crossings between the transitions (Fig. \ref{fig:4K_magnet_sweep}(a-b)) which we clearly also see in the measured data. Using just these equations we can get a remarkable agreement with the data, as shown in Fig. \ref{fig:4K_magnet_sweep}(c). We use the \textsc{QuTiP} Python package \cite{qutip1,qutip2} to calculate the transition energies numerically from the Hamiltonians above.

Notably, we find that the data shows eight transition lines instead of six. The two extra lines are the outermost and less bright lines. These transitions are seemingly violating the transition selection rules $\Delta m=m_h-m_e=-1,0,1$, where $m_h$ and $m_e$ are the hole and electron spins, respectively. However, due to the mixing of the hole states at low fields (close to the avoided crossings) these transitions are not strictly forbidden. As the magnetic field increases the forbidden transition lines are getting fainter, which is the expected result, since the hole states become less mixed and the spin-values of the hole states are getting better defined at higher magnetic fields.

From the data in Fig. \ref{fig:4K_magnet_sweep}(c) we can also extract the g-factors ($g_d,g_1,g_2$) defined in Eqs.~(\ref{eq:donorZeeman}) and (\ref{eq:holezeeman}). The donor electron g-factor $g_d$ anisotropy and strain dependence are well studied and we use the literature value 1.9985 for all data. The hole $g$-factors are known to have strong anisotropy between different crystal orientations \cite{Kaminskil1980,BeckThesis,Litvinenko2016}. Here we determine the hole $g$-factors in [011] orientation, and fits to the data give $g_1$ and $g_2$ parameters shown in Table~\ref{tab:fits}. Our $g_1$ and $g_2$ are well in line with the measurements in other crystal directions and also agree well with the values obtained in \cite{Litvinenko2016} at the high magnetic field quadratic Zeeman regime. Using the measured $g_1$ and $g_2$, we can also extract the g-factors for heavy-holes and light-holes (the slopes of the transitions if there would be no strain induced crossings) $g_{LH}$ and $g_{HH}$. These values are also shown in Table~\ref{tab:fits} with known literature values for several crystal orientations. Details of analytical calculation of $g_{LH}$ and $g_{HH}$ are shown in Section \ref{sec:gfactor}. They can also be extracted numerically.
 
\section{Polarization dependence}
\label{sec:polarization}
In all optical transitions the polarization of the absorbed or emitted light is intimately tied to the spin angular momentum change of the transition. Thus we would also expect a polarization dependence of the transitions, based on the change in the spin quantum number $\Delta m$. As can be seen from the energy level diagram in Fig. \ref{fig:exciton_scheme}(c), there are two transitions with $\Delta m=-1,\, 0,\, 1$ each. To convert these to the expected light polarization, one must also consider the relative direction of the light with regards to the magnetic dipole moment. In our geometry where the magnetic field is perpendicular to the light travel direction (Voigt geometry), we would expect to be able to excite the $\Delta_m = 0$ transitions ($\pi$-transitions) with one linear polarization component and the $\Delta m=\pm 1$ components ($\sigma$-transitions) with the orthogonal linear polarization.

\begin{figure}
    \centering
    \includegraphics[width=.45\textwidth]{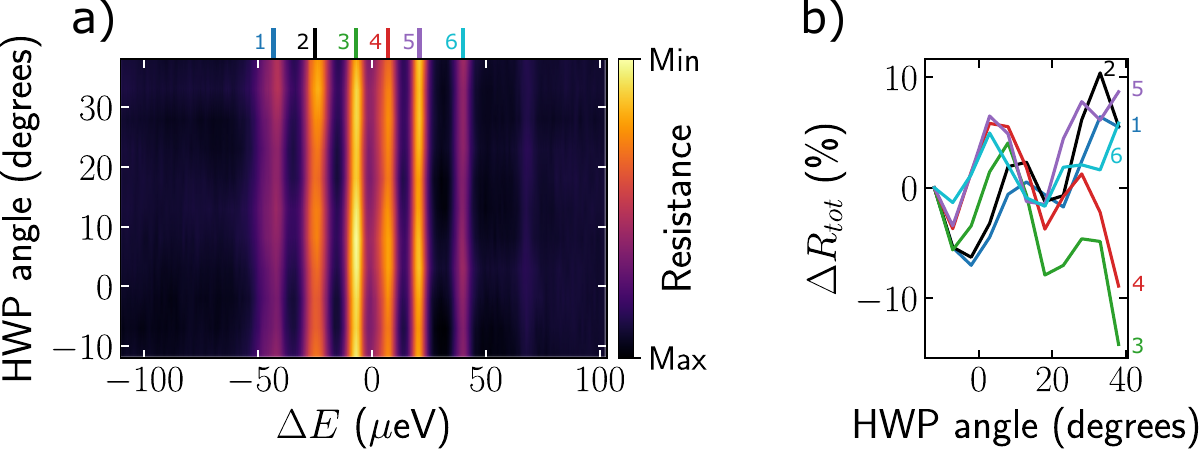}
    \caption{ (a) Half-waveplate (HWP) angle dependence on the exciton signal at 427 mT. (b) Peak height profiles for each of the transitions. $\Delta R_{tot}$ is the change in total resistance. Color labels are given at the top of figure (a). Each line is normalized to start from zero in (b). The absence of proper polarization control is due to hole mixing. There is still a small trend visible where the middle transitions 3 and 4 ($\pi$-transitions in absence of hole mixing) get weaker and other transitions ($\sigma$-transitions in absence of hole mixing) get stronger from left to right as would be expected. This trend is partially obscured by some parasitic polarization dependent power oscillations (the ``fast'' oscillations) in the data.}
    \label{fig:polarizationB}
\end{figure}

The hole-mixing changes this picture as the hole spin states are not anymore well defined and hence also the polarization dependence of the transitions is relaxed. And indeed we see only a very small polarization dependence in our signal as can be seen in Fig. \ref{fig:polarizationB}, where the polarization dependence of the signal is depicted at 427 mT. We see all six allowed transitions at all polarizations even at this field, with only a very slight variation in the amplitude. We also can still distinguish both forbidden transitions (the higher energy one being significantly brighter), the appearance of which shows that there is still a significant amount of hole spin state mixing. (This is also apparent in our numerical modelling based on the parameters above.) 

\section{Above band gap light, laser power dependency and linewidths}
\begin{figure}
    \includegraphics[width=0.45\textwidth]{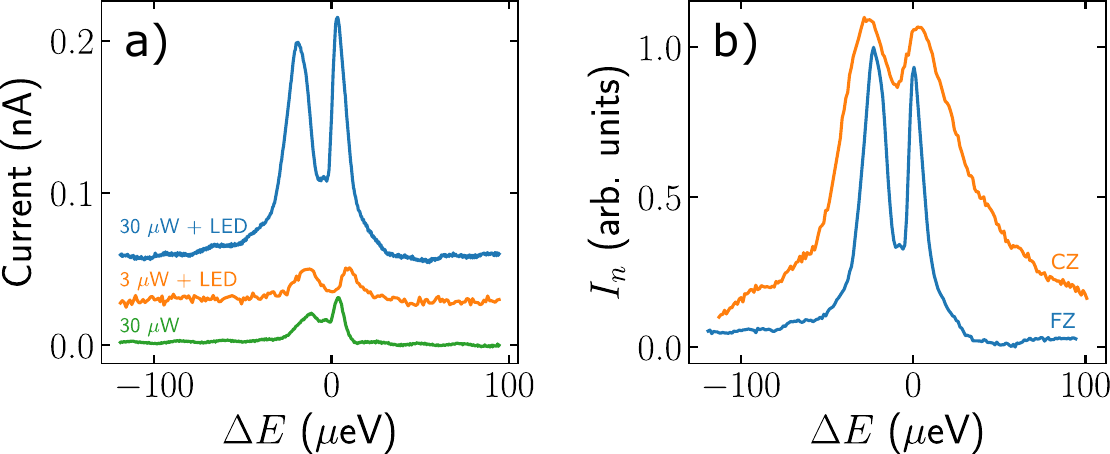}
    \caption{a) Laser power dependence of the exciton signal at zero magnetic field. Current values are offset to show the difference in the peak heights. The base current difference between LED on and LED off measurement is $\approx$3.5~nA. b) Exciton transition at zero applied magnetic field for CZ and FZ silicon. Current values are normalized and CZ data is offset to illustrate the difference in the linewidths. CZ data was captured without LED illumination and FZ data with LEDs. The laser power was the same in both measurements.}
    \label{fig:power dependence}
\end{figure}

We now turn to the more technical details of the hybrid electro-optic readout. First, for good signal-to-noise ratio it seems to be crucial to provide above bandgap light. In Fig. \ref{fig:power dependence}(a) we show the zero magnetic field data measured at different optical powers, with and without the LEDs providing the above bandgap light. It is obvious that the effect of the LEDs is considerable in increasing the signal. This is due to the extra electrons in the conduction band allowing for faster charge neutralization rate of the donors, increasing the "recycling" rate of electrons and hence the measured current, as has also been reported before \cite{Yang2006,Steger2011,Lo2015}.

Interestingly, the shape of the zero-field data changes when the above bandgap light is either on or off. When the LEDs are on the zero-field data does not have a dependency on the laser power, as shown in Fig. \ref{fig:power dependence}(a) showing a constant ZFS of roughly 23~$\mu$eV. However, when the LEDs are off, we see a significant change in the shape of the data. Naively fitting still a two-peak function to the LEDs off data would lead to a ZFS to roughly 15~$\mu$eV. If this would imply a ZFS change due to heating from the LEDs changing strain, the sample would be heating up to 100~K according to \textsc{Comsol} simulations, which is not possible as at these temperatures neither the donor electrons nor excitons would be bound.

Figure \ref{fig:power dependence}(b) shows comparison data at zero applied magnetic field in CZ and FZ samples. As is clearly visible the linewidth improves considerably when moving to FZ-silicon, as is expected from the lower concentration of oxygen and carbon impurities \cite{Safonov1996}. The change in the doping levels might also play a smaller role \cite{Shiraki1979}. In the CZ sample the full-width-half-maximums (FWHM) are roughly 25~$\mu$eV whereas in FZ we see linewidth of 8~$\mu$eV for the heavy-hole (lower energy) peak and 5~$\mu$eV for the light-hole (higher energy) peak. When fitting to data acquired at a high magnetic field we find FWHMs ranging 3--5~$\mu$eV, which is well in line with values reported before for natural silicon \cite{Yang2009}.

\section{Millikelvin measurements}
\label{sec:milliK}
\begin{figure}
    \centering
    \includegraphics[width=.45\textwidth]{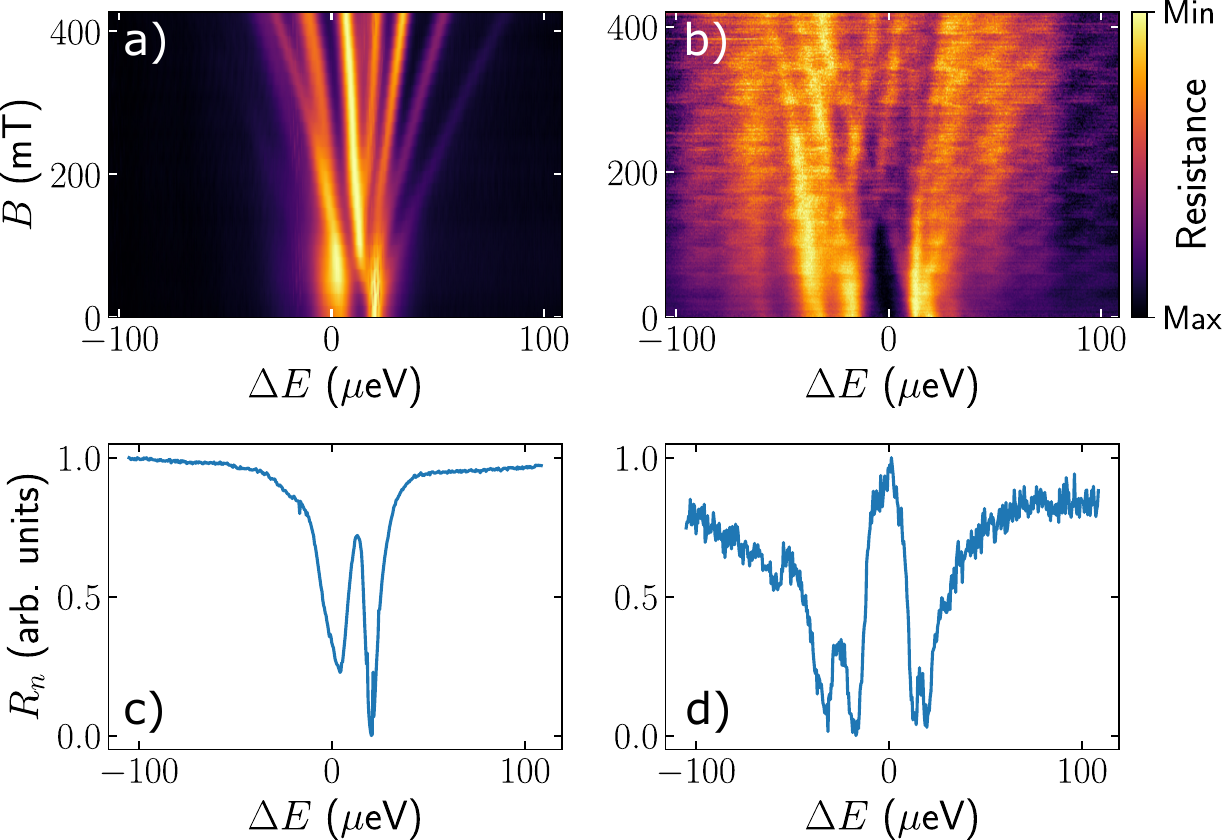}
    \caption{Measured magnetic field dependence on the spectrum at millikelvin temperatures with a) 30~$\mu$W (sample stage temperature 115 mK) and b) 3~$\mu$W of laser power (sample stage temperature 45 mK). Figure (c) and (d) show crosscuts at zero magnetic field. $R_n$ is the normalized resistance.}
    \label{fig:mK data}
\end{figure}

In addition to 3.3 K measurements, we also performed measurements at millikelvin temperatures. However, we did not see a significant difference to the 3.3 K data when high laser power of 30~$\mu$W was used, as shown in Fig. \ref{fig:mK data}(a) and (c) (sample stage thermometer temperature 115 mK). We assume that the heating from the laser power was enough to basically keep the sample at the same temperature as before (it also raised the fridge temperature). When the laser power was reduced to one tenth, the signal became very weak and we observe a very different pattern, see Fig. \ref{fig:mK data}(b) and (d) (sample stage thermometer temperature 45 mK). LEDs providing above bandgap light were on in both measurements. We cannot explain the low power data with strain effects, since the thermal contraction at low temperatures should be negligible. We note that 'additional' zero field splitting has been observed before at 1.4 K temperatures \cite{Yang2006} and was attributed to other charge centers interacting with the spin state. However, the additional splitting in \cite{Yang2006} is one to two orders of magnitude smaller than what we report here.

\section{Discussion and future outlook}
\label{sec:discussion}
We have characterized the phosphorous bound exciton transitions in silicon using hybrid electro-optical readout. The inevitable strain coming from the electrodes needed for this readout method causes hole mixing which both complicates the analysis of the transitions and prevents polarization dependent addressing. Importantly, it does not however fundamentally prevent electron spin readout.

The strain can be seen as a non-wanted property, when dealing with sensitive exciton transitions, but it could also be useful for tuning them. However the tuning needs to take place already in the design of the system. Strain engineering in micro-electro-mechanical-systems is already widely studied topic \cite{Shchepetov2013}, so the methods for the strain tuning are already available. This could be useful for e.g. purposefully matching the transition frequency for photonic components.

We have also directly measured the heavy-hole and light-hole $g$-factors in [011] orientation, and our results confirm the expectation from the $g_1$ and $g_2$ measured at other crystal orientations and at higher fields at this orientation. Indeed, we get a remarkable agreement between theory and experiment for the magnetic field dependency using the Pikus-Bir strain Hamiltonian. It is notable that the $g$-factor anisotropy of holes also allows one to tune the exciton transition position just by rotating the sample, which could be done in-situ with a rotating stage.

For widespread applications, it would be beneficial to integrate the hybrid readout with silicon photonics components. This would require integrating the readout electronics with SOI devices, where strain effects can be increased due to the thin device layer. On the other hand, in a SOI substrate it is easy to suspend the device layer. If one then places the readout electronics on the non-suspended part, the strain caused by the electrodes around the donors is actually \textit{decreased} compared to the bulk case (see strain simulations in Section \ref{sec:strain}). More importantly, with our simulation parameters all the strain components have completely flat profiles in the suspended parts, as we leave 10 $\mu$m space between the electrode and the suspended Si film. This uniformity in the strain field will be crucial for any larger scale architecture in order to avoid inhomogenous broadning of the exciton transition. 

\begin{figure}
    \centering
    \includegraphics[width=.45\textwidth]{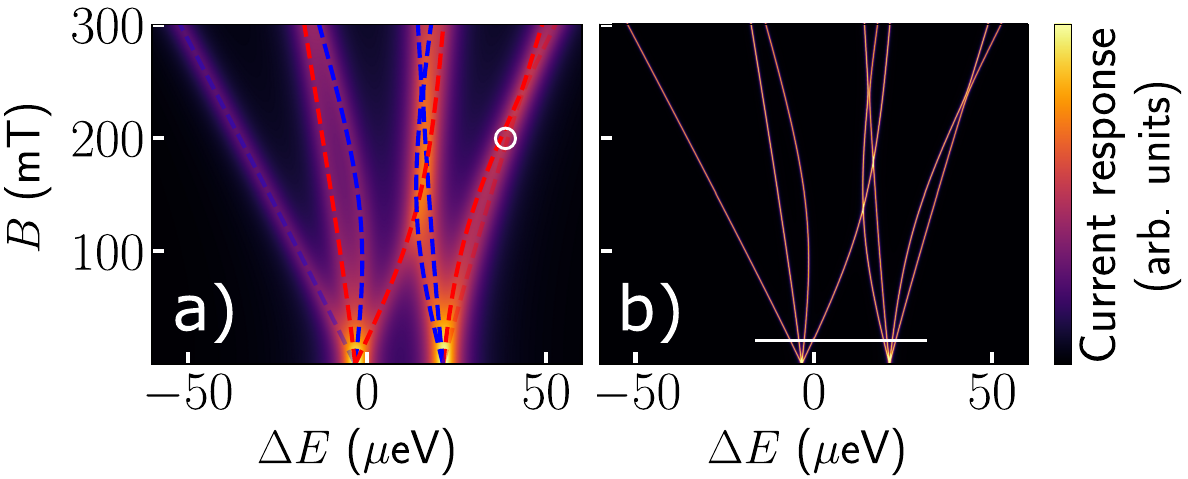}
    \caption{Simulated exciton transitions on SOI platform in a) natural silicon and in b) 99.991\% enriched $^{28}$Si. $B$ is the applied external magnetic field and $\Delta E$ is the transition energy offset from the value without strain or magnetic field. Simulation parameters are presented in Table \ref{tab:fits} and strain is taken from simulations shown in Sec. \ref{sec:strain}. Transition linewidth in a) is the bulk value we have extracted from natural silicon 4~$\mu$eV, and in b) 150~neV as reported in \cite{Yang2006}. The white ring in a) shows possible spin readout conditions in natural silicon at just below 200 mT, where the transition is isolated enough to be addressed optically, and similarly for the white line in (b) at 20 mT.
    }
    \label{fig:SOIspeculation}
\end{figure}

If in addition isotopically pure silicon is used, the ensemble linewidths could be extremely narrow. Previously it has been reported that the exciton transitions in 99.991 \% isotopically enriched $^{28}$Si have a FWHM of 150 neV \cite{Yang2006}. We show expected (simulated) signals from the SOI samples made with natural silicon and enriched $^{28}$Si in Fig. \ref{fig:SOIspeculation} using the measured parameters and strain values from the SOI simulation (assuming strain free substrate at room temperature). From these it is expected that hybrid electro-optical spin readout should be possible in SOI platforms at low magnetic fields.

\section{Methods}
\subsection{Samples}
The CZ wafer used was a 380~$\mu$m thick phosphorous doped wafer acquired from Okmetic Oy with crystal direction (100). The wafer resistivity range at room temperature was specified to be 0.2-0.25~$\Omega\,\text{cm}$, meaning doping levels of 2.4-3.2e16~$\text{cm}^{-3}$.
The FZ wafer used was a 200~$\mu$m thick phosphorous doped wafer acquired from Sil'Tronix ST with crystal direction (100).The wafer resistivity range at room temperature was specified to be 1-5~$\Omega\,\text{cm}$, meaning doping levels of 0.9-4.9e15~$\text{cm}^{-3}$.

The electrodes were fabricated using electron beam lithography with PMMA resist and UHV electron beam evaporation. The electrodes consist of a 5~nm titanium layer for adhesion to silicon surface and 50~nm of gold.

\subsection{Strain simulations}
\label{sec:strain}
The strain in our samples was estimated by simulating the thermal expansion mismatches in \textsc{COMSOL} Multiphysics software. In the simulation a silicon chip rests on a copper stage and has gold electrodes on the top. We assume zero-strain at room temperature and then calculate the induced strain (by both the electrodes and the stage) at low temperatures. We made custom interpolation tables for the linear thermal expansion coefficients \cite{Rodriguez2018,Pamato2018,White1964}, since at low temperatures the coefficients are considerably different compared to the room temperature case. The temperature is swept down to 3.3~K and the strain is calculated for each temperature step. An anisotropic model and parameters were used for silicon.

\begin{figure}
    \centering
    \includegraphics[width=.45\textwidth]{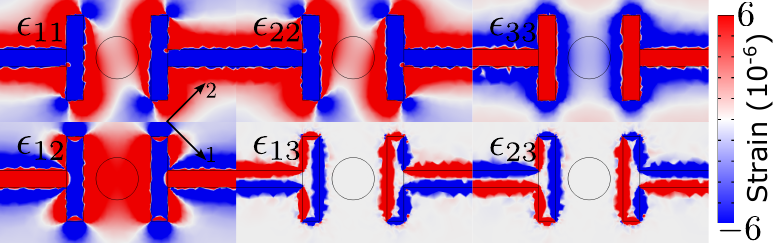}
    \includegraphics[width=.45\textwidth]{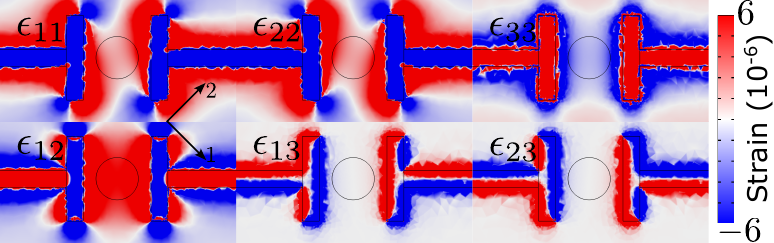}
    \caption{Simulated strain tensor elements on the sample surface (top) and at \textcolor{red}{1}~$\mu$m depth (bottom) in the bulk sample. The black arrows are showing the coordinate directions and the circle between the electrodes is showing the location and size of the laser spot.}
    \label{fig:strainsurf}
\end{figure}

The strain varies considerably as a function of depth and an interesting problem is the question on which depths does our signal originate. It turns out, we can answer this question by considering the inhomogenous broadening caused by the strain and comparing that to our measured transition linewidths. This analysis shows that our signal only originates from the first couple of micrometers of the substrate, otherwise we should see a much more pronounced widening of the transition linewidths according to our model, see Supplementary Material section~\ref{sec:straindetails} for more details. Hence, in Fig.~\ref{fig:strainsurf} we plot all the strain components at the surface and at 1~$\mu$m depth.

\begin{figure}
    \centering
    \includegraphics[width=.45\textwidth]{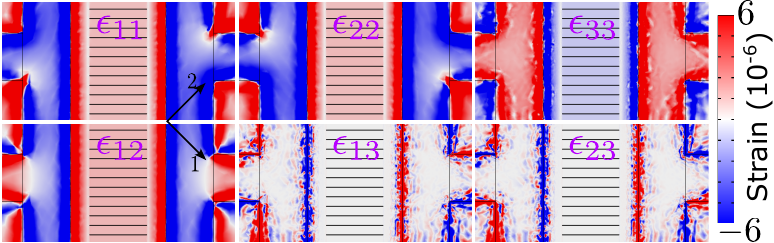}
    \caption{Simulated strain tensor elements at 100~nm depth in the suspended 220 nm silicon film. The black arrows are showing the coordinate directions. The stripes between the electrodes are narrow holes that allow the release of the Si film by HF etching. Note, how constant the strain components are within the released area.}
    \label{fig:strain100nm}
\end{figure}

We repeated the simulation also for SOI material that has a 220 nm natural Si film, 3~$\mu$m SiO$_2$ layer and 750~$\mu$m Si substrate. In the simulations we suspend the area between the electrodes, leaving a $10$~$\mu$m gap between the suspension and the electrode edge. This allows the strain, created by the electrodes, to relax before the suspended silicon film starts and considerably both lessens the strain and relaxes any gradients in the suspended area as can be seen in the simulated strain components at 100 nm depth that are shown in Fig. \ref{fig:strain100nm}.

\subsection{Extracting hole g-factors}
\label{sec:gfactor}
The Zeeman energy splitting of a 3/2-hole can be calculated from \cite{bir1974}
{\small
\begin{align}
    E_{1,2} =& \pm \mu_0 \left\{ \frac{\mathbf{B}^2}{8} \left[ 9 \left( g_1 + \frac{9}{4} g_2 \right)^2 + \left( g_1 + \frac{g_2}{4} \right)^2 \right] + \right.  \nonumber \\
    & \left. \left( g_1 + \frac{7}{4} g_2 \right) \left[ \left( g_1 + \frac{13}{4} g_2 \right)^2 \mathbf{B}^4 - \right.\right. \nonumber \\
    & \left.\left. 9g_2\left( g_1 + \frac{5}{2} g_2 \right) \left( B^2_x B^2_y + B^2_x B^2_z + B^2_y B^2_z \right) \right]^{\frac{1}{2}} \right\}^{\frac{1}{2}}
    \label{eq:EHH}
\end{align}
}%
and
{\small
\begin{align}
    E_{3,4} =& \pm \mu_0 \left\{ \frac{\mathbf{B}^2}{8} \left[ 9 \left( g_1 + \frac{9}{4} g_2 \right)^2 + \left( g_1 + \frac{g_2}{4} \right)^2 \right] - \right.  \nonumber \\
    & \left. \left( g_1 + \frac{7}{4} g_2 \right) \left[ \left( g_1 + \frac{13}{4} g_2 \right)^2 \mathbf{B}^4 - \right.\right. \nonumber \\
    & \left.\left. 9g_2\left( g_1 + \frac{5}{2} g_2 \right) \left( B^2_x B^2_y + B^2_x B^2_z + B^2_y B^2_z \right) \right]^{\frac{1}{2}} \right\}^{\frac{1}{2}}
    \label{eq:ELH}
\end{align}
}%
where $E_{1,2}$ and $E_{3,4}$ give the heavy-hole and light-hole splitting energies, respectively.
By setting the magnetic field direction using notation $[xyz]$, referring to the crystal orientation, one can calculate the effective $g$-factors along any direction. The heavy hole and light hole $g$-factors, that are observed at the high field asymptote, can be solved from
\begin{align}
        &\frac{1}{2}\mu_0 g_{LH} \mathbf{B} =  E_{3,4} \label{eq:gLH}\\
        &\frac{3}{2}\mu_0 g_{HH} \mathbf{B} =  E_{1,2}. \label{eq:gHH}
\end{align}

\begin{acknowledgments}
We acknowledge useful discussions with Jarryd Pla and Joe Salfi, and Henri Lyyra for proofreading the manuscript. This project has received funding from the European Research Council (ERC) under the European Union’s Horizon 2020 research and innovation programme (grant agreement No 852428), from Academy of Finland Grant No 321416 and from Jenny and Antti Wihuri Foundation.
\end{acknowledgments}

\clearpage
\onecolumngrid

\renewcommand{\thefigure}{S\arabic{figure}}
\renewcommand{\thetable}{S\arabic{table}}
\renewcommand{\theequation}{S\arabic{equation}}

\setcounter{figure}{0}
\setcounter{equation}{0}
\setcounter{section}{0}
\setcounter{subsection}{0}

\section*{Supplementary Information}

\subsection{Measurement setup details}
\label{sec:measurementsetup}
The measurements are done in ``dry'' dilution fridge (Bluefors LD400), which can be operated at both millikelvin and 3 K temperatures. The laser (tunable diode laser 1045-1085~nm, Newport TLB-6722) is guided via free space from outside the cryostat to a fixed lens (NA=0.83) above the sample chip. The laser spot diameter is roughly 50~$\mu$m in the data presented, and hence far from the diffraction limit. The spot size is estimated from the infrared camera view. The sample is glued on a piezo stage (Attocube ANPxyz101/2) which allows us to align and focus the laser. Current and voltage are measured over the sample with a low noise current amplifier (Femto DLPCA-200), and a low noise voltage preamplifier (DL instruments model 1201), respectively. The magnetic field is created by a homemade superconducting magnet in the Helmholtz configuration. Hall sensor (Asensor Technology HE244) is used to measure magnetic field during measurements. 

\subsection{Stark shift}
\label{sec:efield}
\begin{figure}
    \centering
    \includegraphics[width=.8\textwidth]{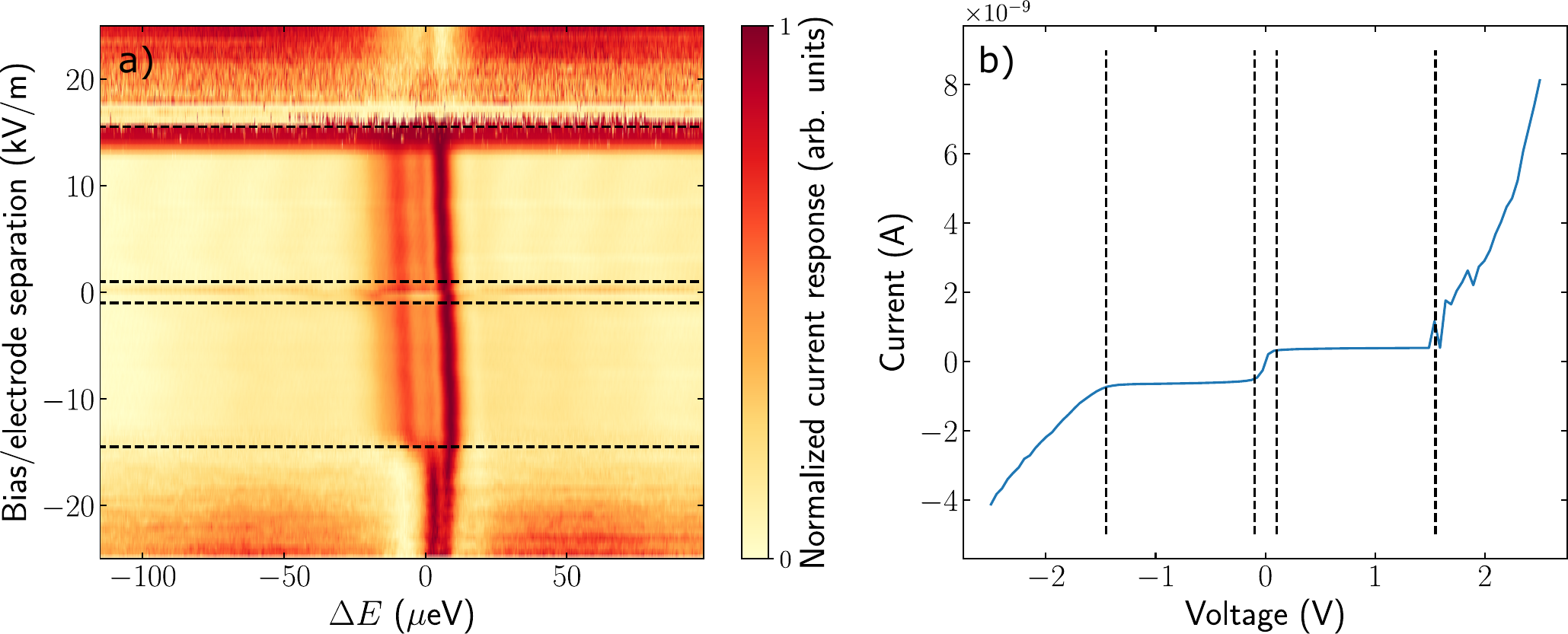}
    \caption{a) Exciton resonance peak as a function of bias voltage in the FZ silicon sample. b) IV-characteristics of the FZ silicon sample when laser is off resonance. Both a) and b) were measured without above band gap light on. Dashed lines in a) match the lines in b). At the flat regions the additional voltage drop is happening at the Schottky barriers and not over the measured sample area.}
    \label{fig:Efieldsweep}
\end{figure}

We also performed measurements sweeping the bias voltage at zero magnetic field in order to characterize the Stark shift in our devices at 3.3~K. Figure \ref{fig:Efieldsweep} shows bias voltage sweep from -2.5~V to 2.5~V. Electrode separation for this FZ sample was 100~$\mu$m. Above band gap light was turned off for this measurement. IV-curve extracted from the data in Fig. \ref{fig:Efieldsweep}(a) is shown in Fig. \ref{fig:Efieldsweep}(b). IV-curve shows clear characteristics of a system with two Schottky diodes in series \cite{Grillo2021}. We can infer from the IV-curve (see modelling below) that the electric field over the sample does not change in regions where the diodes rectify the voltage (from -1.5~V to -0.1~V and from 0.1~V to 1.5~V). These points can be seen clearly in Fig. \ref{fig:Efieldsweep}(a). 
We discuss more about the IV-characteristics in the next section. It is however interesting to note that the data exhibits a feature where the ZFS gets considerably smaller. Hence, we speculate that it could be possible that one could use electric fields to partially overcome the effects of strain induced ZFS.

\subsection{Details of IV-characteristics}
\label{sec:IVdetails}

\begin{figure}
    \centering
    \includegraphics[width=.45\textwidth]{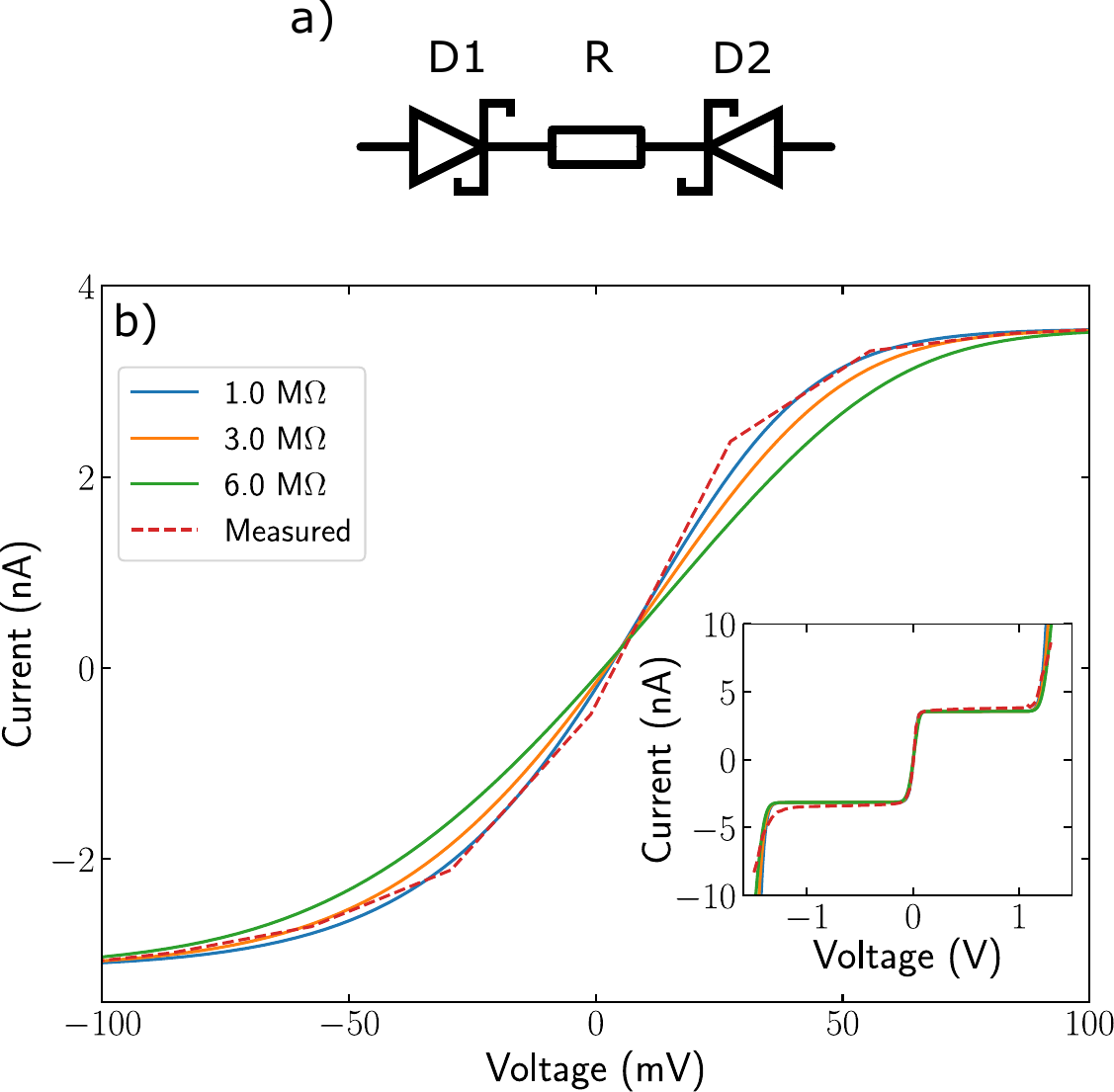}
    \caption{a) Circuit model of the sample. Metal electrodes on semiconductor form two Schottky diodes in series with the sample, which is modeled as a resistor. b) IV-characteristics of the circuit in a) for three different sample resistor values and the IV-curve of the FZ silicon sample with the above bandgap LED light and the laser on. Inset shows the IV-characteristics on a longer scale. Diodes D1 and D2 have slightly different parameters to get a better fit to the data. Parameters for diode D1 were: saturation current $I_S=3.35$ nA, breakdown voltage $B_V=1.1$ V and emission coefficient $N=0.5$. And parameters for diode D2 were: $I_S=3.35$ nA, $B_V=1.3$ V and $N=0.9$.}
    \label{fig:ltspice sim}
\end{figure}

To further understand the Stark shift experiment, we did a network analysis study for the exciton sample. The sample was modeled as two Schottky diodes in series with a resistor. Schottky diodes represent the metal-silicon interfaces and the resistor plays the role of silicon. The circuit used in the analysis is shown in Fig. \ref{fig:ltspice sim}(a). IV-characteristics were simulated using \textsc{LTSpice} software and are shown for three different resistance values in Fig. \ref{fig:ltspice sim}(b). Note that the diode parameters were chosen to get a best fit for the data with 1~M$\Omega$ resistor. Furthermore, there is also small offset in the simulated data to get better fit to the measured data (current $+0.2$~nA and voltage $+5$~mV). However, we were able to get equally good fit with different resistor values just by adjusting the diode parameters. Therefore, we can't say anything about the absolute resistance values of the sample based on these simulations as we do not know the Schottky diode parameters. Nonetheless, there is a linear dependency between measured voltage and resistor values when the current bias is between $-2$~nA and $2$~nA, which is the region where data in Figs. \ref{fig:4K_magnet_sweep}(c), \ref{fig:polarizationB}(a), and \ref{fig:mK data} have been measured.

We noticed major increase in the noise level when the measurement was operated at or above the breakdown voltage (meaning above 1.5 V and below -1.5 V) and increasing the bias even further from the breakdown voltage would completely remove the exciton signal. This increase in noise level can be seen clearly in Fig. \ref{fig:Efieldsweep}a). We speculate that this noise is avalanche noise coming from operating the diode at breakdown voltage.

\subsection{Strain modeling}
\label{sec:straindetails}

\begin{figure}
    \centering
    \includegraphics[width=.75\textwidth]{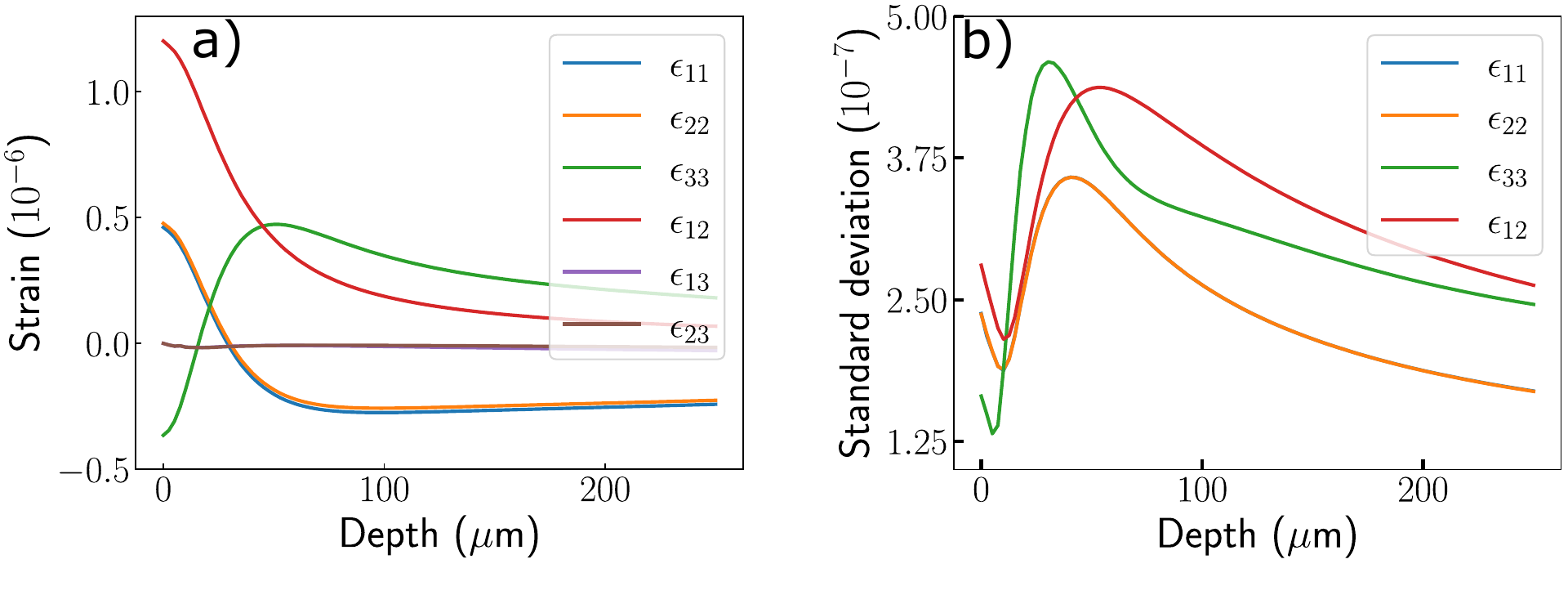}
    \caption{a) The average of the strain within the 50 $\mu$m laser spot as a function of depth. This is the average of all points up to the specified depth. b) The standard deviation of the strain components as a function of depth. Again, this is taking into account all mesh points up to the specified depth.}
    \label{fig:strainZ}
\end{figure}

The model geometry used for the strain model included the whole silicon chip with the electrodes, and a sample stage, where the chip was glued. A thermal expansion mismatch problem was solved using finite element method in \textsc{COMSOL} software. The strain field values from the solved \textsc{COMSOL} model were exported and the statistics of the strain field were studied with a \textsc{Python} code. The code only takes into account strain values inside a cylindrical, 50~$\mu$m laserspot area between the electrodes. The cylinder was sliced into 2.5~$\mu$m thick discs and average values of the strain components were extracted from each disc. In addition, the standard deviation of the strain components was calculated. The data are shown in Fig.~\ref{fig:strainZ}.

Interestingly, we can deduce a limit for the depth from which our signal originates by studying the standard deviation of the strain and hence the implied inhomogenous broadening of the signal. The fact that we do not see considerable broadening of the exciton peaks would seem to imply that all our signal comes from a shallow range, possibly up to 10~$\mu$m depth. If the signal would originate any deeper there should be a considerable broadening of the transitions. In addition, as the average values of the strain components also change as a function of depth, the shape of the avoided crossings in the data is best fitted if we only consider the first 2.5~$\mu$m of the substrate. Based on these, we conclude that the majority of the observed signal comes from within the first 2.5 $\mu$m. We speculate the shallow depth is due to the LEDs above bandgap light not penetrating deeper to the substrate, but we have no comprehensive model for this. According to simulations, there is no significant variation in the strain in the x-y direction, as our laser spot does not overlap with the electrodes as seen in Fig.~\ref{fig:strainsurf}.

In calculating the exciton transition energies from the Pikus-Bir Hamiltonian Eq.~(\ref{eq:HPB}), we used the strain component and standard deviation values that represent the average over the first 2.5 $\mu$m depth: $\epsilon_{11}=(4.6\pm2.4)\cdot10^{-7}$, $\epsilon_{22}=(4.8\pm2.4)\cdot10^{-7}$, $\epsilon_{33}=(-3.5\pm1.6)\cdot10^{-7}$,  $\epsilon_{12}=(-11.8\pm2.8)\cdot10^{-7}$ and $\epsilon_{13}=\epsilon_{23}=0$. With the standard deviation reported here, we estimate the strain induced inhomogeneous broadening to be 2.5 $\mu$eV at most. Finally, we note that the sign of $\epsilon_{12}$ had to be changed compared to \textsc{COMSOL} simulations. A positive sign would lead to a large discrepancy with the data as the sign influences greatly the locations and the magnitudes of the avoided crossings. We assume that the issue has to do with the sign convention of \textsc{COMSOL}, but other option is that correct $d$ should be positive.

\end{document}